\documentclass[runningheads]{llncs}
\usepackage{graphicx}
\usepackage{url}

\usepackage{ragged2e}

\usepackage{amsmath}                               
\usepackage{amssymb}

\usepackage{multirow}

\usepackage{graphics}
\usepackage{url,amsfonts,epsfig,graphicx}
\usepackage{epsf}

\usepackage{subfigure}

\usepackage{comment}

\usepackage{microtype}
\usepackage{xspace,xargs}   

\usepackage[misc,geometry]{ifsym}  

\usepackage{marginnote}  
\newcommand{\todo}[1]{\marginpar{\tiny \flushleft{#1}}}

\def\bigS{\mathcal{S}\xspace}

\def\MS{{\normalfont\textsf{MS}}\xspace}

\def\ourAlgo{{\rm cLCP-mACS}\xspace}

\def\BWT{{\rm BWT}\xspace}
\def\EBWT{{\rm EBWT}\xspace}

\def\LCP{{\rm LCP}\xspace}

\def\ACS{{\rm ACS}\xspace}

\def\cLCP{cLCP\xspace}

\def\ebwt{{\normalfont\textsf{ebwt}}\xspace}

\def\flcp{{\normalfont {\rm LCP}}\xspace} 
\def\anull{{\normalfont\texttt{null}}\xspace}

\def\lcp{{\normalfont\textsf{lcp}}\xspace}

\def\id{{\normalfont\textsf{id}}\xspace}

\def\Uclcp{{\normalfont\textsf{UcLCP}}\xspace}
\def\Lclcp{{\normalfont\textsf{LcLCP}}\xspace}
\def\clcp{{\normalfont\textsf{cLCP}}\xspace}

\def\arrayD{{\normalfont\textsf{D}}\xspace} 
\def\D{{\normalfont\textsf{Score}}\xspace} 

\newcommandx{\todomari}[2][1=]{\todo[linecolor=green,backgroundcolor=green!25,bordercolor=green,#1]{#2}}

\begin{document}
\frontmatter          
\pagestyle{headings}  

\title{The colored longest common prefix array\\computed via sequential scans\thanks{G.R. and M.S. are partially and D.V. is supported by the project Italian MIUR-SIR CMACBioSeq (``Combinatorial methods for analysis and compression of biological sequences'') grant n.~RBSI146R5L.}
}
\titlerunning{The colored longest common prefix array computed via sequential scans}

\newcommand{\repeatthanks}{\textsuperscript{\thefootnote}}

\author{
Fabio Garofalo\inst{1} 
\and
Giovanna Rosone\inst{2}\thanks{Corresponding author}
\and \\
Marinella Sciortino\inst{1}  
\and 
Davide Verzotto\inst{2}   
} 

\authorrunning{F. Garofalo, G. Rosone, M. Sciortino \and D. Verzotto} 

\institute{
University of Palermo, Palermo, Italy\\
\email{garofalo$\_$uni@yahoo.it,$\;$marinella.sciortino@unipa.it}
\and University of Pisa, Pisa, Italy\\
\email{giovanna.rosone@unipi.it,$\;$davide.verzotto@di.unipi.it}
}

\maketitle              

\begin{abstract}
Due to the increased availability of large datasets of biological sequences, the tools for sequence comparison are now relying on efficient alignment-free approaches to a greater extent. Most of the alignment-free approaches require the computation of statistics of the sequences in the dataset. Such computations become impractical in internal memory when very large collections of long sequences are considered.
In this paper, we present a new conceptual data structure, the \emph{colored longest common prefix array} (cLCP), that allows to efficiently tackle several problems with an alignment-free approach. In fact, we show that such a data structure can be  computed via sequential scans in semi-external memory. By using cLCP, we propose an efficient lightweight strategy to solve the \emph{multi-string Average Common Substring (\ACS) problem}, that consists in the pairwise comparison of a single string against a collection of $m$ strings simultaneously, in order to obtain $m$ \ACS induced distances. Experimental results confirm the effectiveness of our approach.

\end{abstract}


\section{Introduction}

The rapid increase in the availability of large sets of biological sequences observed in the last two decades, particularly triggered by the human sequencing project, posed several challenges in the analysis of such data.  
So far, traditional methods based on sequence alignment worked well for small and closely related sequences, but scaling these approaches up to multiple divergent sequences, especially of large genomes and proteomes, is  a difficult task.  
To keep pace with this, several algorithms that go beyond the concept of sequence alignment have been developed, called alignment-free~\cite{Zielezinski2017}. 
Alignment-free approaches have been  explored in several large-scale biological 
applications ranging, for instance, from DNA sequence comparison~\cite{Chang1994,GogMS2010,VerzottoIrredundant,CoxJakobiRosoneST2012,MantaciRRS08} 
to whole-genome phylogeny construction~\cite{ACS2,VerzottoUA,Chor2012,kmacs2014,ALFRED2016} 
and the classification of protein sequences~\cite{VerzottoIrredundant}. 
Most alignment-free approaches above mentioned require, each with its own specific approach and with the use of appropriate data structures, the computation of statistics of the sequences of the analyzed collections. However, it is interesting to note that the increasing number of completely sequenced genomes has caused the computation of many  statistics to be impracticable in internal memory, determining the need for lightweight strategies for the comparative analysis of very large collections of long sequences.

In this paper, we propose a new conceptual data structure, the \emph{colored longest common prefix array} (\clcp), that implicitly stores all the information necessary for computation of statistics on distinguishing, repeating or matching strings within a collection or different collections. Loosely speaking, given a collection $\bigS$, in which each string (or subset of strings) can be identified by a specific color, we can generally define \clcp as an integer array representing the longest common prefix between any specific suffix of a string $s_r \in \bigS$ and the nearest suffix of a specific string $s_t \in \bigS$ in the sorted list of suffixes of $\bigS$. Here, we assume that $\bigS$ is partitioned in two subsets and consider the suffixes of strings belonging to different subsets, but we remark that one can consider any situation and note also that the definition can be easily adapted to more than two sets. We also show that $\clcp$ can be computed via sequential scans and therefore acquires the characteristics of an appropriate structure for analyzing large collections of strings stored in external memory. 

Such a data structure can be used in several applicative contexts. In this paper we explore the multi-string Average Common Substring  (\ACS)~\cite{ACS2} problem. 
More specifically, the \ACS measure is a simple and effective 
 alignment-free method for pairwise string comparison~\cite{Chor2012,ALFRED2016},
based on the concept of \emph{matching statistics} (\MS)~\cite{Chang1994,Gusfield97,GogMS2010}.
Given two strings $s$ and $t$, it can be defined as a couple of arrays $\MS(s,t)$ and $\MS(t,s)$ that store, at each position $i$, the length of the longest substring that starts at position $i$ of the string given as first parameter that is also a substring of the string given as second parameter.

\ACS\ approach has been employed in several biological applications ~\cite{ACS2,VerzottoUnic,kmacs2014,VerzottoUA,VerzottoBeyond}. Generalization of measures based on longest matches with mismatches have been proposed in \cite{ApostolicoGP14}, also with distributed approaches \cite{PetrilloGP17}.
Similarly to~\cite{SunReinert2013}, we define the \emph{multi-string \ACS problem} as the pairwise comparison of a single string, say $s_\chi \in \bigS^0$ of length $n_\chi$, against a set of $m$ strings, say $s_r \in \bigS^1$ with $1 \leq r \leq m$, by considering 
the strings in $\bigS^1$ all together, in order to obtain $m$ \ACS induced distances at once.
A major bottleneck in the computation (and application)  of \ACS and \MS initially consisted in the construction of a suffix tree. More recent approaches use efficient indexing structures \cite{BelazzouguiCunial_SPIRE2014}, CDAWG \cite{BelazzouguiCunial_SPIRE2017}, backward search~\cite{GogMS2010} or enhanced suffix arrays~\cite{kmacs2014}.
However, to our knowledge, the above mentioned approaches would require a great effort, especially in terms of RAM space, if applied to very large collections of long strings. 

In this paper we use $\clcp$ to efficiently solve the above mentioned multi-string \ACS problem. Preliminary experimental results show that our algorithm is a competitive tool for the lightweight simultaneous computation of a pairwise distance of a string versus each string in a collection, allowing us to suppose that this data structure and its computational strategy can be used for more general versions of the multi-string \ACS problem.



	\section{Preliminaries}\label{Sec_definitions}
	
	Let $\Sigma =\{c_1, c_2, \ldots, c_\sigma\}$ be a finite ordered alphabet with $c_1< c_2< \ldots < c_\sigma$, where $<$ denotes the standard lexicographic order.
	We consider finite strings such as $s\in \Sigma^*$, where $s[1], s[2],\ldots,s[n]$ denote its characters and $|s| = n$ its length. 
	A \emph{substring} of a string $s$ is written as $s[i,j] = s[i] \cdots s[j]$, with a substring $s[1,j]$ being called a \emph{prefix}, while a substring $s[i,n]$ is referred to as a \emph{suffix}.	 A range is delimited by a square bracket if the correspondent endpoint is included, whereas a parenthesis means that the endpoint 
	is excluded.


    The \BWT \cite{bwt94} is a well known reversible string transformation widely used in data compression.
The \BWT can be extended to a collection of strings $\bigS=\{s_1,s_2,\ldots,s_{m}\}$. 
Such an extension, known as \EBWT or multi-string \BWT, is a reversible transformation that produces a string (denoted by $\ebwt(\bigS)$) that is a permutation of the characters of all strings in $\bigS$ \cite{MantaciRRS07}.  Lightweight implementations of \EBWT have been proposed \cite{BauerCoxRosoneTCS2013,Louza2017,EgidiLouzaManziniTelles2018_arxiv}.
We append to each string $s_i$ of length $n_i$ a distinct end-marker symbol $\$_i < c_1$ (for implementation purposes, we could simply use a unique end-marker $\$$ for all strings in $\bigS$). The output string $\ebwt(\bigS)$ is the concatenation of the symbols (circularly) preceding each suffix of the strings in $\bigS$, sorted according to the lexicographic order. More in detail, the length of $\ebwt(\bigS)$ is denoted by $N=\sum_{i=1}^{m}n_i + m$ and $\ebwt(\bigS)[i]=x$, with $1 \leq i \leq N$, if $x$ circularly precedes the $i$-th suffix $s_j[k,n_j+1]$ (for some $1\leq j\leq m$ and $1\leq k\leq n_j+1$), according to the lexicographic sorting of the suffixes of all strings in $\bigS$. In this case we say the suffix $s_j[k,n_j+1]$ is associated with the position $i$ in $\ebwt(\bigS)$.
We can associate to each string $s_i\in \bigS$ a color $i$ in $ID= \{1, 2, \dots, m\}$. The output string $\ebwt(\bigS)$, enhanced with the array $\id(\bigS)$ of length $N$ where $\id(\bigS)[i]=r$, with $1\leq r\leq m$ and $1 \leq i \leq N$, if $\ebwt(\bigS)[i]$ is a symbol of the string $s_r \in \bigS$, is called \emph{colored \EBWT}. 

The \emph{longest common prefix} (\LCP) array of the collection $\bigS$~\cite{PuglisiTurpin2008,CGRS_JDA_2016,Louza2017} is the array $\lcp(\bigS)$ of length $N+1$,  such that $\lcp(\bigS)[i]$, with $2 \leq i \leq N$, is the length of the longest common prefix between the suffixes associated to the positions $i$ and $i-1$ in $\ebwt(\bigS)$ and $\lcp(\bigS)[1] = \lcp(\bigS)[N+1]= -1$ set by default. 
We denote by $\flcp(i, j)$
the length of the \LCP
 between the suffixes associated with positions $i$ and $j$ in $\ebwt(\bigS)$, i.e.\
 $\min \{\lcp(\bigS)[l]: i < l \leq j\}$.  
An interval $[i, j]$ with  $1 \leq i < j \leq N$, is an \emph{k-lcp interval} if $\lcp(\bigS)[i] < k$, $\flcp(i, j)=k$, $\lcp(\bigS)[j + 1]<k$.
The set $\bigS$ will be later omitted if it is clear from the context.

\section{Colored Longest Common Prefix Array}\label{sec:clcp}

In this section we present a novel data structure, the  \emph{colored longest common prefix array} (\clcp).  
Loosely speaking, the \clcp\ array represents the longest common prefix between a suffix that belongs to a string of the collection $\bigS$  and the nearest suffix belonging to another string of $\bigS$, in the list of sorted suffixes of $\bigS$.
In this paper,  for simplicity of description, we assume that $\bigS$ is partitioned in two subsets and consider the suffixes of strings belonging to different subsets, but we remark that one can consider any situation and note also that the definition can be easily adapted for more than two sets.




For $i=1, \ldots, N$ and $t=1,\ldots m$, let $prev(i,t) = \max \{x\!\mid\!1\!\leq\!x<i, \id(\bigS)[x]\!=\!t\}$ and $next(i,t) = \min \{x\!\mid\!i<\!x\!\leq\!N, \id(\bigS)[x]\!=\!t\}$ (if such an $x$ exists, and \anull otherwise).


In order to give the notion of the $\clcp$ array, we first define the {\em Upper colored} \LCP\ array ($\Uclcp$) and the {\em Lower colored} \LCP\ array ($\Lclcp$), as follows.	
\begin{definition}\label{Def_cLCPup}
The \emph{upper} (resp. \emph{lower}) \emph{colored longest common prefix} array ($\Uclcp$) (resp. $\Lclcp$) is a ($N \times m$)-integer array where, for each $i_r \in\{ 1,2,\ldots,N\}$ with $id[i_r]=r$ and $t \in ID$, 
$\Uclcp[i_r][t] = \flcp(prev(i_r,t), i_r)$ (resp. $\Lclcp[i_r][t]=\flcp(i_r, next(i_r,t))$).  Both $\flcp (\anull, i_r)$ and $\flcp (i_r, \anull)$ are set equal to $0$.        
\end{definition}

	\begin{definition}\label{Def_cLCP}
The \emph{colored longest common prefix} array ($\clcp$) is a ($N \times m$)-integer array where, for each $i_r \in\{ 1,2,\ldots,N\}$ with $id[i_r]=r$ and $t \in ID$, 
$\clcp[i_r][t] = \max(\Uclcp[i_r][t], \Lclcp[i_r][t]) $. 
	\end{definition}

\begin{table}[!t]
\centering
{ 
\scriptsize
	$$
	\begin{array}{|@{\ }c@{\ }|c@{\ }|c@{\ }|c@{\ }|c@{\ }|c@{\ }|c@{}|c@{\ }|c@{\ }|c@{\ }|c@{\ }|c@{\ }|c@{\ }|c@{\ }|c@{\ }|c@{\ }|c@{\ }|c@{\ }|c@{\ }c@{\ }c@{\ }c@{\ }c@{\ }c@{\ }c@{\ }c@{\ }c@{\ }c@{\ }c@{\ }c@{\ }c@{\ }c@{\ }c@{\ }|}
		\hline 
		\multicolumn{9}{|l|}{} & \multicolumn{3}{c|}{}           & \multicolumn{3}{c|}{}           & \multicolumn{3}{c|}{}            &                                &                                 &                                 &                                 &                                 &            &            &            &            &           &           &           &           &           &           \\[-0.175cm]
		\multicolumn{9}{|l|}{} & \multicolumn{3}{c|}{\Uclcp}           & \multicolumn{3}{c|}{\,\Lclcp}           & \multicolumn{3}{c|}{\,\clcp}            &                                &                                 &                                 &                                 &                                 &            &            &            &            &           &           &           &           &           &           \\
		\hline
					&          &        &  &          &           &         &           &           &          &           &           &          &           &           &          &           &           &  \multicolumn{15}{l|}{} \\[-0.21cm]
		\#			& {\,\id}         & {\,\bigS}        & {\,\ebwt} & {\,\lcp}         & {\,\arrayD}          & \lcp_{\chi}        & {\,\alpha}          & {\,\zeta}          & {\,\chi}         & {\,1}          & {\,2}          & {\,\chi}         & {\,1}          & {\,2}          & {\,\chi}         & {\,1}          & {\,2}          &  \multicolumn{15}{l|}{\mbox{ Sorted suffixes of  $\bigS$}} \\
		\hline
		\textbf{1}  & \mathbf{\chi} & \textbf{1} & \textbf{C} & \textbf{-1} & \textbf{0} & \textbf{-1} & \mathbf{\infty} & \textbf{0} & \textbf{} & \textbf{0} & \textbf{0} & \textbf{} & \textbf{0} & \textbf{0} & \textbf{} & \textbf{0} & \textbf{0} & \textbf{}                      & \mathbf{\$_\chi}                      & \textbf{}                      & \textbf{}                      & \textbf{}                      & \textbf{} & \textbf{} & \textbf{} & \textbf{} & \textbf{} & \textbf{} & \textbf{} & \textbf{} & \textbf{} &           \\
		2           & 1          & 0          & T & 0           & 0          &             & 0          & 0          & 0         &            &            & 0         &            &            & 0         &            &            &                                & \$_1                               &                                &                                &                                &           &           &           &           &          &          &          &           &           &         \\
		3           & 2          & 0          & A & 0           & 0          &             & 0          & 0          & 0         &            &            & 0         &            &            & 0         &            &            &                                & \$_2                               &                                &                                &                                &           &           &           &           &          &          &          &          &          &          \\ \cline{20-20}
		4           & 2          & 0          & C & 0           & 2          &             & 0          & 1          & 0         &            &            & 1         &            &            & 1         &            &            & \multicolumn{1}{c|}{}          & \multicolumn{1}{c|}{A}          & \$_2                               &                                 &                                 &            &            &            &            &           &           &           &           &           &         \\
		5           & 2          & 0          & \$_2 & 1           & 0          &             & 0          & 1          & 0         &            &            & 1         &            &            & 1         &            &            & \multicolumn{1}{c|}{}          & \multicolumn{1}{c|}{A}          & A                               & C                               & G                               & C          & C          & G          & C          & C         & G         & G         & C         & A         & \$_2         \\ \cline{20-22}
		6           & 1          & 0          & \$_1 & 1           & 4          &             & 0          & 3          & 0         &            &            & 3         &            &            & 3         &            &            & \multicolumn{1}{c|}{}          & \multicolumn{1}{c|}{A}          & C                               & \multicolumn{1}{c|}{G}          & A                               & G          & A          & C          & G          & A         & T         & \$_1         &           &           &          \\
		7           & 1          & 0          & G & 4           & 0          &             & 0          & 3          & 0         &            &            & 3         &            &            & 3         &            &            & \multicolumn{1}{c|}{}          & \multicolumn{1}{c|}{A}          & C                               & \multicolumn{1}{c|}{G}          & A                               & T          & \$_1          &            &            &           &           &           &           &           &          \\ \cline{20-23}
		8           & 2          & 0          & A & 3           & 5          &             & 0          & 4          & 0         &            &            & 4         &            &            & 4         &            &            & \multicolumn{1}{c|}{}          & \multicolumn{1}{c|}{A}          & C                               & \multicolumn{1}{c|}{G}          & \multicolumn{1}{c|}{C}          & C          & G          & C          & C          & G         & G         & C         & A         & \$_2         &            \\
		\textbf{9}  & \mathbf{\chi} & \textbf{1} & \mathbf{\$_0} & \textbf{4}  & \textbf{0} & \textbf{0}  & \mathbf{\infty} & \textbf{0} & \textbf{} & \textbf{3} & \textbf{4} & \textbf{} & \textbf{1} & \textbf{0} & \textbf{} & \textbf{3} & \textbf{4} & \multicolumn{1}{c|}{\textbf{}} & \multicolumn{1}{c|}{\textbf{A}} & \textbf{C}                      & \multicolumn{1}{c|}{\textbf{G}} & \multicolumn{1}{c|}{\textbf{C}} & \textbf{G} & \textbf{C} & \textbf{C} & \mathbf{\$_\chi} & \textbf{} & \textbf{} & \textbf{} & \textbf{} & \textbf{} &            \\ \cline{20-23}
		10          & 1          & 0          & G & 1           & 0          &             & 1          & 0          & 1         &            &            & 0         &            &            & 1         &            &            & \multicolumn{1}{c|}{}          & \multicolumn{1}{c|}{A}          & G                               & A                               & C                               & G          & A          & T          & \$_1          &           &           &           &           &           &           \\
		11          & 1          & 0          & G & 1           & 0          &             & 1          & 0          & 1         &            &            & 0         &            &            & 1         &            &            & \multicolumn{1}{c|}{}          & \multicolumn{1}{c|}{A}          & T                               & \$_1                               &                                 &            &            &            &            &           &           &           &           &           &          \\ \cline{20-20}
		\textbf{12} & \mathbf{\chi} & \textbf{1} & \textbf{C} & \textbf{0}  & \textbf{2} & \textbf{0}  & \mathbf{\infty} & \textbf{0} & \textbf{} & \textbf{0} & \textbf{0} & \textbf{} & \textbf{1} & \textbf{1} & \textbf{} & \textbf{1} & \textbf{1} & \multicolumn{1}{c|}{\textbf{}} & \multicolumn{1}{c|}{\textbf{C}} & \mathbf{\$_\chi}                      & \textbf{}                      & \textbf{}                      & \textbf{} & \textbf{} & \textbf{} & \textbf{} & \textbf{} & \textbf{} & \textbf{} & \textbf{} & \textbf{} & \textbf{}  \\
		13          & 2          & 0          & G & 1           & 0          &             & 1          & 0          & 1         &            &            & 1         &            &            & 1         &            &            & \multicolumn{1}{c|}{}          & \multicolumn{1}{c|}{C}          & A                               & \$_2                               &                                 &            &            &            &            &           &           &           &           &           &          \\ \cline{20-21}
		\textbf{14} & \mathbf{\chi} & \textbf{1} & \textbf{G} & \textbf{1}  & \textbf{3} & \textbf{1}  & \mathbf{\infty} & \textbf{0} & \textbf{} & \textbf{0} & \textbf{1} & \textbf{} & \textbf{1} & \textbf{2} & \textbf{} & \textbf{1} & \textbf{2} & \multicolumn{1}{c|}{\textbf{}} & \multicolumn{1}{c|}{\textbf{C}} & \multicolumn{1}{c|}{\textbf{C}} & \mathbf{\$_\chi}                      & \textbf{}                      & \textbf{} & \textbf{} & \textbf{} & \textbf{} & \textbf{} & \textbf{} & \textbf{} & \textbf{} & \textbf{} & \textbf{}  \\
		15          & 2          & 0          & G & 2           & 0          &             & 2          & 0          & 2         &            &            & 1         &            &            & 2         &            &            & \multicolumn{1}{c|}{}          & \multicolumn{1}{c|}{C}          & \multicolumn{1}{c|}{C}          & G                               & C                               & C          & G          & G          & C          & A         & \$_2         &           &           &           &           \\
		16          & 2          & 0          & G & 3           & 0          &             & 2          & 0          & 2         &            &            & 1         &            &            & 2         &            &            & \multicolumn{1}{c|}{}          & \multicolumn{1}{c|}{C}          & \multicolumn{1}{c|}{C}          & G                               & G                               & C          & A          & \$_2          &            &           &           &           &           &           &           \\ \cline{20-21}
		17          & 1          & 0          & A & 1           & 3          &             & 1          & 2          & 1         &            &            & 2         &            &            & 2         &            &            & \multicolumn{1}{c|}{}          & \multicolumn{1}{c|}{C}          & \multicolumn{1}{c|}{G}          & A                               & G                               & A          & C          & G          & A          & T         & \$_1         &           &           &           &          \\
		18          & 1          & 0          & A & 3           & 0          &             & 1          & 2          & 1         &            &            & 2         &            &            & 2         &            &            & \multicolumn{1}{c|}{}          & \multicolumn{1}{c|}{C}          & \multicolumn{1}{c|}{G}          & A                               & T                               & \$_1          &            &            &            &           &           &           &           &           &          \\ \cline{20-23}
		\textbf{19} & \mathbf{\chi} & \textbf{1} & \textbf{G} & \textbf{2}  & \textbf{5} & \textbf{1}  & \mathbf{\infty} & \textbf{0} & \textbf{} & \textbf{2} & \textbf{1} & \textbf{} & \textbf{0} & \textbf{4} & \textbf{} & \textbf{2} & \textbf{4} & \multicolumn{1}{c|}{\textbf{}} & \multicolumn{1}{c|}{\textbf{C}} & \multicolumn{1}{c|}{\textbf{G}} & \multicolumn{1}{c|}{\textbf{C}} & \multicolumn{1}{c|}{\textbf{C}} & \mathbf{\$_\chi} & \textbf{}  & \textbf{}  & \textbf{}  & \textbf{} & \textbf{} & \textbf{} & \textbf{} & \textbf{} & \textbf{}  \\
		20          & 2          & 0          & A & 4           & 0          &             & 4          & 0          & 4         &            &            & 3         &            &            & 4         &            &            & \multicolumn{1}{c|}{}          & \multicolumn{1}{c|}{C}          & \multicolumn{1}{c|}{G}          & \multicolumn{1}{c|}{C}          & \multicolumn{1}{c|}{C}          & G          & C          & C          & G          & G         & C         & A         & \$_2         &           &           \\
		21          & 2          & 0          & C & 5           & 0          &             & 4          & 0          & 4         &            &            & 3         &            &            & 4         &            &            & \multicolumn{1}{c|}{}          & \multicolumn{1}{c|}{C}          & \multicolumn{1}{c|}{G}          & \multicolumn{1}{c|}{C}          & \multicolumn{1}{c|}{C}          & G          & G          & C          & A          & \$_2         &           &           &           &           &           \\ \cline{20-23}
		\textbf{22} & \mathbf{\chi} & \textbf{1} & \textbf{A} & \textbf{3}  & \textbf{0} & \textbf{3}  & \mathbf{\infty} & \textbf{0} & \textbf{} & \textbf{2} & \textbf{3} & \textbf{} & \textbf{0} & \textbf{2} & \textbf{} & \textbf{2} & \textbf{3} & \multicolumn{1}{c|}{\textbf{}} & \multicolumn{1}{c|}{\textbf{C}} & \multicolumn{1}{c|}{\textbf{G}} & \multicolumn{1}{c|}{\textbf{C}} & \textbf{G}                      & \textbf{C} & \textbf{C} & \mathbf{\$_\chi} & \textbf{} & \textbf{} & \textbf{} & \textbf{} & \textbf{} & \textbf{} & \textbf{} \\ \cline{20-22}
		23          & 2          & 0          & C & 2           & 0          &             & 2          & 0          & 2         &            &            & 0         &            &            & 2         &            &            & \multicolumn{1}{c|}{}          & \multicolumn{1}{c|}{C}          & \multicolumn{1}{c|}{G}          & G                               & C                               & A          & \$_2          &            &            &           &           &           &           &           &           \\ \cline{20-21}
		24          & 1          & 0          & A & 0           & 2          &             & 0          & 1          & 0         &            &            & 1         &            &            & 1         &            &            & \multicolumn{1}{c|}{}          & \multicolumn{1}{c|}{G}          & A                               & C                               & G                               & A          & T          & \$_1          &            &           &           &           &           &           &           \\
		25          & 1          & 0          & C & 2           & 0          &             & 0          & 1          & 0         &            &            & 1         &            &            & 1         &            &            & \multicolumn{1}{c|}{}          & \multicolumn{1}{c|}{G}          & A                               & G                               & A                               & C          & G          & A          & T          & \$_1         &           &           &           &           &            \\
		26          & 1          & 0          & C & 2           & 0          &             & 0          & 1          & 0         &            &            & 1         &            &            & 1         &            &            & \multicolumn{1}{c|}{}          & \multicolumn{1}{c|}{G}          & A                               & T                               & \$_1                               &            &            &            &            &           &           &           &           &           &           \\ \cline{20-21}
		27          & 2          & 0          & G & 1           & 3          &             & 0          & 2          & 0         &            &            & 2         &            &            & 2         &            &            & \multicolumn{1}{c|}{}          & \multicolumn{1}{c|}{G}          & \multicolumn{1}{c|}{C}          & A                               & \$_2                               &            &            &            &            &           &           &           &           &           &           \\ \cline{20-22}
		\textbf{28} & \mathbf{\chi} & \textbf{1} & \textbf{C} & \textbf{2}  & \textbf{4} & \textbf{0}  & \mathbf{\infty} & \textbf{0} & \textbf{} & \textbf{1} & \textbf{2} & \textbf{} & \textbf{0} & \textbf{3} & \textbf{} & \textbf{1} & \textbf{3} & \multicolumn{1}{c|}{\textbf{}} & \multicolumn{1}{c|}{\textbf{G}} & \multicolumn{1}{c|}{\textbf{C}} & \multicolumn{1}{c|}{\textbf{C}} & \mathbf{\$_\chi}                      & \textbf{}  & \textbf{}  & \textbf{}  & \textbf{}  & \textbf{} & \textbf{} & \textbf{} & \textbf{} & \textbf{} & \textbf{} \\
		29          & 2          & 0          & C & 3           & 0          &             & 3          & 2          & 3         &            &            & 2         &            &            & 3         &            &            & \multicolumn{1}{c|}{}          & \multicolumn{1}{c|}{G}          & \multicolumn{1}{c|}{C}          & \multicolumn{1}{c|}{C}          & G                               & C          & C          & G          & G          & C         & A         & \$_2         &           &           &           \\
		30          & 2          & 0          & C & 4           & 0          &             & 3          & 3          & 3         &            &            & 2         &            &            & 3         &            &            & \multicolumn{1}{c|}{}          & \multicolumn{1}{c|}{G}          & \multicolumn{1}{c|}{C}          & \multicolumn{1}{c|}{C}          & G                               & G          & C          & A          & \$_2          &           &           &           &           &           &          \\ \cline{20-22}
		\textbf{31} & \mathbf{\chi} & \textbf{1} & \textbf{C} & \textbf{2}  & \textbf{0} & \textbf{2}  & \mathbf{\infty} & \textbf{0} & \textbf{} & \textbf{1} & \textbf{2} & \textbf{} & \textbf{0} & \textbf{1} & \textbf{} & \textbf{1} & \textbf{2} & \multicolumn{1}{c|}{\textbf{}} & \multicolumn{1}{c|}{\textbf{G}} & \multicolumn{1}{c|}{\textbf{C}} & \textbf{G}                      & \textbf{C}                      & \textbf{C} & \mathbf{\$_\chi} & \textbf{} & \textbf{} & \textbf{} & \textbf{} & \textbf{} & \textbf{} & \textbf{} & \textbf{} \\ \cline{20-21}
		32          & 2          & 0          & C & 1           & 0          &             & 1          & 0          & 1         &            &            & 0         &            &            & 1         &            &            & \multicolumn{1}{c|}{}          & \multicolumn{1}{c|}{G}          & G                               & C                               & A                               & \$_2          &            &            &            &           &           &           &           &           &           \\ \cline{20-20}
		33          & 1          & 0          & A & 0           & 0          &             & 0          & 0          & 0         &            &            & 0         &            &            & 0         &            &            &                                & T                               & \$_1                               &                                 &                                 &            &            &            &            &           &           &           &           &           &           \\
		&            &            &  & -1          &            & -1          &            &            &           &            &            &           &            &            &           &            &            &                                &                                 &                                 &                                 &                                 &            &            &            &            &           &           &           &           &           &           \\ 
		\hline
	\end{array}
	$$
}		
		\caption{ 
			 Let $\Sigma$=$\{A,C,G,T\}$, $s_\chi$=$ACGCGCC\$_\chi \in \bigS^0$, $s_1$=$ACGAGACGAT\$_1 \in \bigS^1$, and $s_2$=$AACGCCGCCGGCA\$_2 \in \bigS^1$. Then, $\D(s_\chi,s_1)$=$11/7$, $\D(s_\chi,s_2)$=$19/7$, $\D(s_1,s_\chi)$=$15/10$, $\D(s_2,s_\chi)$=$30/13$, and thus $\ACS(s_\chi,s_1)$=$0.67$ and $\ACS(s_\chi, s_2)$=$0.34$. 
			   In bold are all positions associated with 
			 suffixes of $s_\chi$ (i.e.\ the limits of the $\chi$-intervals). 
		}
		\label{Tab_biscuix_output}
	\end{table}	

For simplicity $\Uclcp$, $\Lclcp$, and $\clcp$ are also defined when $r=t$. For all $i_r$ such that $id[i_r]=r$, $\Uclcp[i_r][t]$ coincides with the correspondent value in the usual $\lcp(\{s_r\})$. 
As mentioned before, note that the notion of $\Uclcp$, $\Lclcp$, and $\clcp$ can be also given for a pair of disjoint collections of strings $\bigS^0$ and $\bigS^1$ by obtaining an array defined for the pairs $(i_r,t)$ with $id[i_r]=r$ and $t \in ID$ such that $s_r$ and $s_t$ belong to a different collection.

A given string $s_\chi\in \bigS^0$ having color $\chi$ implicitly induces a partition of $\lcp(\bigS)$ in open intervals delimited by consecutive suffixes having color $\chi$ (or the position $1$ and $N+1$ of $\lcp$),
called \emph{$\chi$-intervals}. Let us consider a position $i_r$ contained within a $\chi$-interval such that $\id[i_r]=r$ and $s_r\in \bigS^1$.
	  Then, we can use a similar procedure as the one employed 
	in \cite{kmacs2014}, such that 
	\begin{equation}\label{Eq_uclcp_r}
	\Uclcp[i_r][\chi] = \flcp (prev(i_r,\chi), i_r) = \min \{\lcp[x] : prev(i_r,\chi) < x \le i_r\},
	\end{equation}
	\begin{equation}\label{Eq_lclcp_r}
	\Lclcp[i_r][\chi] = \flcp (i_r, next(i_r,\chi)) = \min \{\lcp[x] : i_r < x \le next(i_r,\chi)\}. 
	\end{equation}

Additionally,  we notice that there exists a relationship between the values of $\Uclcp$ related to the suffixes of $s_r$ and the values of $\Lclcp$ related to the suffixes of $s_\chi$. Indeed, if $j_\chi$ is a position where $\id[j_\chi]=\chi$, then 
\begin{equation}\label{Eq_lclcp_chi}
\Lclcp[j_\chi][r] = \flcp(j_\chi, next(j_\chi,r)) = \Uclcp[next(j_\chi,r)][\chi].
\end{equation} 
Similarly, there exists a relationship between the values of $\Uclcp$ related to suffixes of $s_\chi$ and the values of $\Lclcp$ related to suffixes of $s_r$.
In particular, 
	\begin{equation}\label{Eq_uclcp_chi}
	\Uclcp[j_\chi][r] = \flcp(prev(j_\chi,r), j_\chi) = \Lclcp[prev(j_\chi,r)][\chi].
	\end{equation} 

Table~\ref{Tab_biscuix_output} shows the values of \Uclcp, \Lclcp and \clcp of the running example, in which the collection $\bigS$ is partitioned into two subsets $\bigS^0=\{ACGCGCC\$_\chi\}$ and $\bigS^1=\{ACGAGACGAT\$_1, AACGCCGCCGGCA\$_2\}$.

\section{Lightweight computation of the cLCP}
\label{sec:ligthclcp}
	
In this section we describe a lightweight strategy to compute the colored longest common prefix array $\cLCP$. For sake of simplicity we consider the case in which the collection $\bigS$ is partitioned into two subsets  $\bigS^0$ and $\bigS^1$, and $\bigS^0$ consists of a single string $s_\chi$ of length $n_\chi$. The general case can be treated analogously.



\begin{definition}\label{Def_relevant_block}
A \emph{colored} $k$-\emph{lcp interval} is a $k$-lcp interval $[i,j]$ such that, among all the suffixes associated to the range $[i,j]$, at least one suffix belongs to $\bigS^0$ and at least one suffix belongs to $\bigS^1$.
	\end{definition}    
    
    

\begin{definition}\label{Def_PropertyB}
Let $\arrayD[1, N+1]$ denote an integer array, such that $\arrayD[i] = k$ if a colored $(k-1)$-lcp interval starts at position $i$ and for every colored $h$-lcp interval starting at position $i$ then $h\leq k-1$. 

\end{definition}	


Table~\ref{Tab_biscuix_output} highlights the conceptual blocks of suffixes that are associated to the positions $i$ of $\arrayD$ such that $\arrayD[i] \neq 0$.

Note that the array $\arrayD$ can be easily computed in $\Theta(N)$ time by linearly scanning the arrays  $\lcp(\bigS)$ and $\id(\bigS)$ and by using a stack that simulates the computation of the colored $k$-lcp intervals. During the sequential scan, each element can be inserted or deleted from the stack once, at most. 
Furthermore, considering that each suffix could take part in no more than $\max\lcp(\bigS)$ nested blocks, the stack requires $O(\max\lcp(\bigS))$ space, at most.
We note that this upper bound in space is unlikely to be reached in practice, especially since the stack is emptied when two consecutive values of $\id$  corresponding to strings of different subsets are found. It is important to specify that the above mentioned stack could be stored in external memory.

In the following we describe a sequential strategy to construct $\clcp$ of the collection $\bigS$ by using $\id(\bigS)$, $\lcp(\bigS)$, and $\arrayD(\bigS)$. 

Without loss of generality, let us consider a generic string $s_r \in \bigS^1$ and $s_\chi \in \bigS^0$.
Assume that $\ebwt[i_r]$, with $1 \leq i_r \leq N$, is associated with a suffix of $s_r$, i.e.\ $\id[i_r]=r \neq \chi$, and let
$\chi_1 = prev(i_r,\chi)$ 
and
 $\chi_2 = next(i_r,\chi)$. 
Moreover, for simplicity,  
let $\Uclcp_r$ (resp.\ $\Lclcp_r$) denote $\Uclcp$ of $s_r$ versus $s_\chi$ (resp.\ $\Lclcp$ of $s_r$ versus $s_\chi$), i.e.\ the values $\Uclcp[i_r][\chi]$ (resp.\ $\Lclcp[i_r][\chi]$) for all such $i_r$; 
 and $\Lclcp_\chi$ (resp.\ $\Uclcp_\chi$) denote $\Lclcp$ (resp.\ $\Uclcp$) of $s_\chi$ versus $s_r$, i.e.\ the values $\Lclcp[j_\chi][r]$ (resp.\ $\Uclcp[j_\chi][r]$) for all $1 \leq j_\chi \leq N$ such that $\id[j_\chi]=\chi$. 

\paragraph{$\Uclcp_{r}$ computation ---}
This is the easiest case, since Equation~\ref{Eq_uclcp_r} allows us to directly compute $\Uclcp_r$ sequentially and linearly in the total size $N$ of $\lcp$. This enables us to scan $\lcp$ forward only once for all suffixes of all $m$ strings in $\bigS^1$, by keeping track of 
the minimum value found since the beginning of each conceptual $\chi$-interval (see column $\alpha$ in Table~\ref{Tab_biscuix_output}). 
If we consider the $\chi$-interval $(\chi_1, \chi_2)$, 
by employing a variable $\alpha$ we can iteratively compute the minimum value among consecutive elements of $\lcp$ and determine, for every $i_r \in (\chi_1, \chi_2)$, the \LCP between the suffix associated with position $i_r$ and the suffix associated with position $\chi_1$:  
$\Uclcp[i_r][\chi] = \flcp(\chi_1, i_r) = \min\{\lcp[x] : x \in (\chi_1, i_r]\} = \alpha$.
\begin{example} [Running example]
If the $\chi$-interval is $(14, 19)$ and $i_r = 18$, then $\Uclcp[18][\chi] = \flcp(14, 18) = \min\{\lcp[x] : x \in (14, 18]\} = \alpha[18] = 1.$ 
\end{example}

\paragraph{$\Lclcp_{\chi}$ computation ---}
Since $\Lclcp_\chi$ is strictly related to $\Uclcp_r$ by Equation~\ref{Eq_lclcp_chi},
we would like to compute it sequentially and linearly as well.
Suppose that we have just computed $\Uclcp[i_r][\chi]$ and $i_r$ represents the first suffix of $s_r$ encountered since the beginning in $(\chi_1,\chi_2)$. 
Then, by Equation~\ref{Eq_lclcp_chi}, $\Lclcp[\chi_1][r] = \Uclcp[i_r][\chi]$.      
To keep track of the first 
instance of 
every $s_r \in \bigS^1$ in the interval, we could resort to a bit-array 
 of $m$ elements for $\chi_1$.

Nevertheless, this is not sufficient to complete the construction of $\Lclcp_\chi$, because there might be no suffixes of a particular string $s_{r} \in \bigS^1$ within $(\chi_1, \chi_2)$, but other suffixes of $s_{r}$ might exist 
 at positions $>$$\chi_2$. To tackle this issue, once we have thoroughly read $\lcp$ 
 and filled $\Lclcp_{\chi}$ 
 using the above procedure,
 we can propagate the computed values of $\Lclcp_{\chi}$ backward from lower to higher 
  lexicographically ranked suffixes of $\chi$, in order to 
  complete $\Lclcp_{\chi}$. 
 For example, to propagate the information from $\chi_2$ to $\chi_1$, we must compute:
 \begin{equation}\label{Eq_propagation}
 \Lclcp[\chi_1][r] = \min\{\flcp(\chi_1,\chi_2), \Lclcp[\chi_2][r]\}.
 \end{equation}
 Thus, iteratively, we can propagate the information backward from the lowest ranked suffixes of $\chi$ to the top of $\Lclcp_{\chi}$.
 \begin{example} [Running example]  After 
  the first scan of \lcp in the example of Table~\ref{Tab_biscuix_output}, $\Lclcp[12][1]$ (i.e.\ suffix of $s_\chi$ in row 12 versus string $s_1 \in \bigS^1$) would be 0, whereas by propagating the information back from the suffix of $s_\chi$ in row 14, we obtain:   
$ \Lclcp[12][1] = \min\{\flcp(12,14), \Lclcp[14][1]\} 
 =\min\{1, 2\} = 1$.
 \end{example}
 
 

\paragraph{$\Lclcp_{r}$ computation ---}The most interesting  
part 
is computing $\Lclcp_r$ in such a way as to 
avoid the backward scan of $\id$ and $\lcp$ suggested by Equation~\ref{Eq_lclcp_r} and, concomitantly,
reduce the memory footprint required for keeping both $\Uclcp_r$ and $\Lclcp_r$ to a somehow negligible one.
Thus, 
 we show how 
to sequentially determine, for every $i_r \in (\chi_1, \chi_2)$, the \LCP between the suffix associated with position $i_r$ and the suffix associated with position $\chi_2$.

Let us consider the array \arrayD introduced in Definition~\ref{Def_PropertyB}.
Intuitively, \arrayD provides an interlacing forward information 
that could be exploited to compute $\Lclcp[i_r][\chi]$ sequentially, as soon as we reach 
position $i_r$. 
 Firstly, observe that, for any $1 \le i_r \le N$ with $\id[i_r] = r$ and any $\chi_1 < x < \chi_2$, $prev(x,\chi) = prev(i_r,\chi) = \chi_1$ and $next(x,\chi) = next(i_r,\chi) = \chi_2$.

\begin{remark}\label{Rem_N1N2}
For any $x_1 < x_2$, with $\chi_1 \leq x_1 < \chi_2$ and $\chi_1 < x_2 \leq \chi_2$, $\flcp(x_1, x_2) \geq \flcp(\chi_1, \chi_2)$.
\end{remark}


\begin{lemma}\label{Lem_B_array}
For any $1 \le i_r \le N$, if $\flcp(i_r, \chi_2) = k - 1$ then
there exists an $x$, with $\chi_1 < x \le i_r$, such that $\arrayD[x] = k \neq 0$ if and only if $\flcp(\chi_1, \chi_2) < k - 1$.
\end{lemma}

Moreover, it follows that $\arrayD[x]$ would be (the only) maximum in the range $(\chi_1, i_r]$ and its value is $\geq 2$. 
Hence, we can determine $\Lclcp[i_r][\chi]= \flcp(i_r, \chi_2)$.  

\begin{theorem}\label{Th_LcLCP_r}
 \sloppy For any $1 \leq i_r \leq N$ such that $id[i_r] = r$,
	if $\flcp(\chi_1, i_r) > 
	\flcp(\chi_1, \chi_2)$ then 
	$\flcp(i_r, \chi_2) =
	\flcp(\chi_1, \chi_2)$, otherwise $\flcp(i_r, \chi_2) = \max\{\max\{\arrayD[x] : \chi_1 < x \le i_r\} 
	- 1, 
	\flcp(\chi_1, \chi_2)\}$.
\end{theorem}

By using Theorem~\ref{Th_LcLCP_r} we need to keep track of the maximum value (decreased by $1$) among consecutive \arrayD values since the beginning of each conceptual $\chi$-interval (see column $\zeta$ in Table~\ref{Tab_biscuix_output}). An immediate example 
of Theorem~\ref{Th_LcLCP_r}
 is given in column $\Lclcp[\cdot][\chi]$ of Table~\ref{Tab_biscuix_output}, which  
provides the final values of $\Lclcp_r$, 
 where $\flcp(\chi_1, i_r)$ is 
 computed 
 using Equation~\ref{Eq_uclcp_r}
 and $\flcp(\chi_1, \chi_2)$ through $\lcp(\bigS^0)$ (or, shortly, $\lcp_\chi$).
 
 
 \begin{example}[Running example]
\sloppy Let $i_r=16$ (with $prev(i_r,\chi) = 14$ and $next(i_r,\chi) = 19$) such that $\flcp(14, 16) = \Uclcp[16][\chi] = 2 > \flcp(14, 19) = \lcp_\chi[5] = 1$; then,  $\Lclcp[17][\chi] = \flcp(16, 19) = \flcp(14, 19) = 1$. 
 Conversely, by considering $i_r=17$ (with $prev(i_r,\chi) = 14$ and $next(i_r,\chi) = 19$, as before), $\flcp(14, 17) = \Uclcp[17][\chi] = 1 = \flcp(14, 19) = \lcp_\chi[5] = 1$; therefore, 
  $\Lclcp[17][\chi] = \flcp(17, 19) = \max\{\max\{\arrayD[x]: 14 < x \le 17\} - 1,
 \flcp(14, 19)\} = \max\{2, 1\} = 2$.
 Furthermore, we consider the third case of Theorem~\ref{Th_LcLCP_r} such that, for $i_r = 13$ (where $prev(i_r,\chi) = 12$ and $next(i_r,\chi) = 14$), $\flcp(12, 13) = \Uclcp[13][\chi] = 1 = \flcp(12, 14) = \lcp_\chi[4] = 1$ and thus $\Lclcp[13][\chi] = \flcp(13, 14) =  \max\{\max\{\arrayD[x]: 12 < x \le 13\} - 1, \flcp(12, 14)\} = \max\{-1, 1\}=1$.
 \end{example} 
 
 
\paragraph{$\Uclcp_{\chi}$ computation ---}Similarly to $\Lclcp_{\chi}$, 
 we can compute $\Uclcp_{\chi}$ by exploiting Equation~\ref{Eq_uclcp_chi} and the previously computed $\Lclcp_{r}$ within each $\chi$-interval 
  (compare columns $\Uclcp[\cdot][1]$ and $\Uclcp[\cdot][2]$ against column $\Lclcp[\cdot][\chi]$ in Table~\ref{Tab_biscuix_output}).  
To complete the construction of $\Uclcp_\chi$, we need then to propagate forward the information from higher to lower lexicographically ranked suffixes of $\chi$. For example, to propagate the information from $\chi_1$ to $\chi_2$, we must compute
$\Uclcp[\chi_2][r] = \min\{\flcp(\chi_1,\chi_2), \Uclcp[\chi_1][r]\}$.

To reduce the memory footprint, we can use a single matrix 
$\clcp_\chi[1,n_\chi][1,m]$ (initialized with all $0$s) to keep track 
of the maximum values between the corresponding positions of $\Uclcp_\chi$ 
and $\Lclcp_\chi$, which could be 
then refined at most twice by propagation.
Observe that $\Uclcp_\chi$, alone, can be
directly 
computed sequentially, 
eventually reducing 
the additional space to a negligible one of size $O(m)$,
as seen before for $\Uclcp_r$ and $\Lclcp_r$. 

\begin{example}[Running example]
After the first scan of \lcp, 
$\Uclcp[22][1]$ (i.e.\ suffix of $s_\chi$ in row 22 versus string $s_1 \in \bigS^1$) would be $0$, whereas by propagating the information forward from the suffix of $s_\chi$ at row 19, we obtain:   
$\Uclcp[22][1] = \min\{\flcp(19,22), \Uclcp[19][1]\} =\min\{3, 2\} = 2$.
\end{example}



 

 \paragraph{Computational complexity ---}
The first phase of the algorithm consists of the computation of the $\arrayD$ in $\Theta(N)$ time and $O(\max\lcp(\bigS))$. Notice that $\Uclcp_r$ and $\Lclcp_r$ can be determined 
 sequentially (forward) requiring nothing but to update variables $\alpha$ and $\zeta$ keeping track respectively of the minimum among consecutive \lcp values and of the maximum among consecutive \arrayD values since the last $s_\chi$ suffix encountered. 
Moreover one can observe that also in $\Uclcp_\chi$ and $\Lclcp_\chi$ computation both $\lcp_\chi$ and $\clcp_\chi$ are actually accessed either sequentially forward or sequentially backward, up to one position before or after the currently processed one, allowing them to reside in external memory too. This means that we need $O(m)$ additional space in RAM. In order to optimally use the available size $M$ of RAM, assuming $Q \geq 2$ is the number of $m$-elements rows of $\clcp_\chi$ that we could accommodate in RAM,
  at any moment we could just
   keep in memory and process only a single block of $\lcp_\chi$ and $\clcp_{\chi}$ of size proportional to $Q$. Such a block, together with the bit-array of size $m$ required in first part of $\Lclcp_\chi$ computation, yield $O(mQ+\max \lcp(\bigS))$ overall  required space (with $Q$ a configurable parameter). Further, since $\clcp_{\chi}$ values could be refined at most twice by propagation, a global cost of $O(N + mn_{\chi})$ time is deduced.
  Note that, instead, 
  a straightforward approach 
  that just uses Equations \ref{Eq_uclcp_r} and \ref{Eq_lclcp_r} would have required to
   process in RAM at least three data structures, each of size $\sim$$N$, 
   using
   $O(n_\chi N)$ time (without propagation). 
In order to evaluate the $I/O$ operations, we denote by $B$ the disk block size and we assume that both the RAM size and $B$ are measured in units of ($\log N$)-bit words. The overall complexity of the algorithm, included the $I/O$ operations, is synthesized in the following theorem. The number of $I/O$ disk operations is due the fact that the files storing the values of $\id(\bigS)$, $\lcp(\bigS)$, $\arrayD(\bigS)$, $\lcp_{\chi}$ and $\clcp_{\chi}$ are used.


 \begin{theorem}

Let $\bigS$ a collection of $m$ strings. Given $\id(\bigS)[1,N]$, $\lcp(\bigS)[1,N+1]$ and $\lcp(s_\chi)[1,n_\chi+1]$, $\clcp(\bigS)$ can be computed by sequential scans in $\mathcal{O}(N + m n_\chi)$ time and $\mathcal{O}(m+L_1)$ additional space, where $L_1=\max\lcp(\bigS)$. 
The total number of $I/O$ disk operations is  
$O\left(\frac{1}{B \log N}(
N \log m + N \log L_1 +
n_\chi \log L_2 + n_\chi m \log L_1)\right)$,  
where $L_2=\max\lcp(s_\chi)$.
\end{theorem}
%


	
\section{Multi-string ACS Computation by $\cLCP$}
\label{sec:acs}

The \cLCP is a data structure that implicitly stores information useful to compute distinguishing and repeating strings in different collections. Its lightweight computation described in previous section enables the use of $\clcp$ in several contexts in which large collections of long strings are considered. 

Here, we describe its use in the case of computing the matching statistics \MS and the average common substring \ACS measure.     The Average common substring (\ACS) induced distance is based on the 
	concept of \emph{matching statistics} (\MS)~\cite{Chang1994,Gusfield97} 
	and is typically computed by proceeding in two steps.
	Let us first consider two strings $s_r$, of length $n_r$, and $s_t$, of length $n_t$, over $\Sigma$ of size $\sigma$.
		In the first step, \ACS asymmetrically computes the longest match lengths of $s_r$ versus $s_t$, 
	$\MS(s_r, s_t)$, 
	where $s_r$ is the base string.
$\MS(s_r, s_t)[1,n_r]$ is an integer array such that, for any position $j_r$ of $s_r$, $\MS(s_r, s_t)[j_r]$ is the length of the longest prefix of the suffix of $s_r$ starting at position $j_r$ 
that is also a substring 
of $s_t$ (see Table~\ref{tab:tabACS}). 
	 In the second step, \ACS takes the average of these scores 
	$\D(s_r, s_t) = \frac{\sum_{j_r=1}^{n_r} \MS(s_r, s_t)[j_r]}{n_r}$;
	normalizes it by the lengths of $s_r$, $s_t$, and $\sigma$ 
	$Norm(\D(s_r, s_t)) = \frac{{\log}_{\sigma}n_t}{{\D}(s_r, s_t)} - \frac{2\ {\log}_{\sigma}n_r}{(n_r + 1)}$;
	and finally makes the measure symmetrical by defining 
	$\ACS(s_r, s_t) = \frac{Norm(\D(s_r, s_t)) + Norm(\D(s_t, s_r))}{2}$,
	in order to achieve an induced distance.  We observe that \ACS\ is not a metric, because the triangular inequality might not hold in general.  
	Nevertheless, 
	if we assume 
	 $s_r$ and $s_t$ 
	 be 
	generated by finite-state Markovian probability distributions, 
	it follows that \ACS\ is a natural distance measure between these distributions~\cite{ACS2}. 

	\begin{table}[t]
		\centering{
		{\scriptsize	
			$$
			\begin{array}{l} 
			$$
			\begin{array}{|r|cccccccccc|} 
				\hline
				{s_0[j_0]} & {A} & {C} & {G} & {C} & {G} & {C} & {C} & & & {\quad \ \ \;\:\,\!} \\
				{\MS(s_0, s_1)[j_0]} & {3} & {2} & {1} & {2} & {1} & {1} & {1} &  &  &  \\
				\hline
			\end{array}
			$$  \\ 
			
			$$
			\begin{array}{|r|cccccccccc|} 
				\hline
				{s_1[j_1]} & {A} & {C} & {G} & {A} & {G} & {A} & {C} & {G} & {A} & {T} \\
				{\MS(s_1, s_0)[j_1]} & {3} & {2} & {1} & {1} & {1} & {3} & {2} & {1} & {1} & {0} \\
				\hline
			\end{array}
			$$
			\end{array}
			$$
		}
		}
		\caption{Matching statistics $\MS(s_0, s_1)$ and $\MS(s_1, s_0)$ for $s_0$=$ACGCGCC\$_0$ and $s_1$=$ACGAGACGAT\$_1$ on $\Sigma=\{A,C,G,T\}$. It follows that $\D(s_0, s_1)=11/7$, $\D(s_1, s_0)=15/10$ and, thus, $\ACS(s_0, s_1)=0.67$.
		}\label{tab:tabACS}
	\end{table}



For simplicity we assume that we have a set consisting of only one string $\bigS^0 = \{s_{\chi}\}$, of length $n_{\chi}$, and a set of strings $\bigS^1 = \{s_1, s_2, \dots,$ $s_{m}\}$, of length $N^1 = \sum_{1 \leq r \leq m} n_{r}$, and we want to compute the pairwise \ACS\ induced distances between $\bigS^0$ (or, more explicitly, $s_{\chi}$) and every other string in $\bigS^1$ simultaneously, 
 as in the multi-string \ACS problem. Our approach could be also applied to a more general case.  

There is a clear correspondence between \clcp of $s_{\chi}$ versus all strings in $\bigS^1$, and \MS\ of \ACS.
More precisely:
\begin{proposition}\label{Lem_permutation}
Given any two strings $s_r,s_\chi \in \bigS$, $\MS(s_r, s_\chi)$ is a permutation of all values in $\clcp(\bigS)$ related to the suffixes of  $s_r$ (the base string) versus $s_\chi$:
$\MS(s_r, s_\chi) [j_r] = \clcp[i_r][\chi]$,
where $1 \leq i_r \leq N$ such that $\id(\bigS)[i_r]=r$, and $j_r$ is the starting position in $s_r$ of the suffix associated with $\ebwt(\bigS)[i_r]$.
\end{proposition}

Indeed, for each suffix of every string $s_r \in \bigS^1$, associated with $\ebwt[i_r]$, 
$\clcp[i_r]$ would account for the longest prefix 
that is a substring of $s_{\chi}$, and this must correspond to one of the closest suffixes belonging to $s_{\chi}$ immediately above ($prev(i_r,\chi)$) or below ($next(i_r,\chi)$)  row $i_r$ in the sorted suffixes list. 

We can thus exploit the above proposition to compute \MS using \clcp, by using the strategy described in previous section. 
In fact, computing \MS\ by straightly using the Equations \ref{Eq_uclcp_r} and \ref{Eq_lclcp_r} would require to explicitly keep track of \clcp for each $\chi$-interval, which could have width $\Theta(N)$ in the worst case. In this section we show that this additional space can be controlled and reduced by using our lightweight computation of $\clcp$.


Using the construction described in Section~\ref{sec:ligthclcp} we can determine $\Uclcp_r$ and $\Lclcp_r$ sequentially (forward) and these values are definitive (they are not subject to refinement by propagation). Therefore we can reduce multi-string \ACS memory footprint by summing up all the maximum values between the respective positions in $\Uclcp_r$ and $\Lclcp_r$ for every specific string $s_r \in \bigS^1$, and for every position $i_r$, and storing them into an array $\D_{r}$ of size $m$ as they are computed during forward phase, without explicitly maintaining $\clcp_r$ values in internal or external memory. On the other hand, since $\Uclcp_\chi$ and $\Lclcp_\chi$ require propagation to be completed, we need to maintain (a $Q$-sized portion of) $\clcp_\chi$ matrix and similarly cumulate $\clcp_\chi$ values for every position $j_\chi$ and for every string $s_r \in \bigS^1$ into an array $\D_{\chi}$ of size $m$ as these values became definitive during backward phase. Accordingly, multi-string \ACS computation does not add to $\clcp$ construction more than $\Theta(m)$ space and $O(mn_\chi)$ time. Note that in a typical application, $m$ can be assumed $\ll$$N$ and negligible compared to the internal memory available.
Here, we show a simplified version of our strategy described in Section \ref{sec:acs}.
For simplicity, we index the files as array, but the reader can note that we only access to them sequentially.
We need to keep in memory the length of the strings for the $m$ ACS scores.

\section{Preliminary Experiments}


As a proof-of-concept, we test our new data structure (\cLCP) with a prototype tool named
\ourAlgo \cite{cLcpMultiACStool}
designed to solve the multi-string ACS problem. 
All tests were done on a DELL PowerEdge R630 machine, 
$24$-core machine with Intel(R) Xeon(R) CPU E5-2620 v3 at $2.40$ GHz, with $128$ GB of shared memory, used in not exclusive mode. The system is Ubuntu 14.04.2 LTS. 

We show that our preliminary experiments confirm the effectiveness of our approach for the multi-string ACS problem, that consists in the pairwise comparison of a single string against a set of $m$ strings simultaneously, in order to obtain $m$ ACS induced distances.
This is not a limitation, because the computation of pairwise distances between strings of a collection $\bigS$ can be treated analogously, in the sense that one could execute our tool more times, without computing the data structures of the preprocessing step.

We experimentally observed that the preprocessing step is more computationally expensive than the step for computing the $m$ ACS distances via \cLCP.
The problem of computing the $\ebwt(\bigS), \lcp(\bigS), \id(\bigS)$ are being intensively studied, and improving its efficiency is out of the aim of this paper.
So, we omit time/space requirements of the preprocessing step, since (i) these data structures can be reused and (ii) different programs \cite{BEETLTool,extLCPtool,egsaTool,eGAPTool} are used to construct them. So, we solely focus on the phase of computation of the matrix distances.

 

To assess the performance of our algorithm we consider the two collections of genomes listed in \cite{cLcpMultiACStool} and described in Table \ref{table:collection}. 
\begin{table}[t!]
\centering
\begin{tabular}{||c|c|c|c|c||c|c|c||}
\hline
  & $|\ebwt(\bigS)|$ & Min length   & Max length & Max lcp   & Program    & Wall clock  & Memory  \\
  & (Gbytes)  &  	          & 	    	&              &            & (mm:ss)     & (kbytes)  \\
  \hline
1 & 3.434     & 1,080,084  & 10,657,107 & 1,711,194    & \ourAlgo &  27:42  &  6,160 \\
  &                 &            &            &         & kmacs    &  18:50 &    3,723,728    \\
\hline
2 & 9.258  & 744       & 14,782,125   &  5,714,157   &   \ourAlgo &  53:03     &    6,372           \\
  &		 &          &               &                &  kmacs     &  58:08     &   14,717,332    \\
\hline
\end{tabular}
\caption{The first collection contains $932$ genomes, 
the second one contains $4,983$ genomes.
Note that $|s_\chi|=5,650,368$ for the first collection and $|s_\chi|=3,571,103$ for the second one. In both cases these values are greater than the average length of the strings in the respective collection.
The amount of time elapsed from the start to the completion of the instance.
The column memory is the peak Resident Set Size (RSS). Both values were taken with \texttt{/usr/bin/time} command.
}
\label{table:collection}
\end{table}

The auxiliary disk space for the first collection is $34$ Gbytes and $108$ Gbytes for the second one.
But, note that the $\arrayD$ file is sparse, in the sense that it contains many zero-values. 
One can reduce its size by storing only non-zero values or using a different encoding of these values, for instance Sadakane's \cite{Sadakane:2007} encoding. However, this could slow down the computation.

Notice that an entirely like-for-like comparison between our implementation and the above existing implementation is not possible, since, for the best of our knowledge, our tool is the first lightweight tool. 

The ACS computation of two strings has been faced in \cite{ACS2} and it has been implemented by using the $k$ Mismatch Average Common Substring Approach (kmacs) tool \cite{kmacsTool}.
For $k=0$ kmacs exactly computes the ACS. Other algorithms besides kmacs \cite{ALFRED2016,Pizzi16} have been designed to compute alignment-free distances based on longest matches with mismatches, but for the special case $k=0$ kmacs is the software that has the better change to scale with the dataset size. 
The implementation of kmacs works in internal memory and not in sequential way. More in detail, it works in $m$ steps, at each step it builds the suffix array \cite{Manber:1990} and the $\lcp$ array of two strings $s_i$ and $s_j$ (for $1 \leq i < j \leq m$) in order to compute the ACS distance between $s_i$ and $s_j$. 
In our experiments we have fixed $s_\chi = s_1$, so we compare $s_\chi$ with $s_j$ (for $2 \leq j \leq m$). 
Note that the performance in terms of time of kmacs could be improved by separately considering the computation of the auxiliary data structures. However, the occupation of RAM would remain almost the same. 

The experimental results in Table \ref{table:collection} show that our algorithm is a competitive tool for the lightweight simultaneous computation of a pairwise distance of a string versus each string in a collection, in order to obtain an efficient solution for the multi-string ACS problem.


\section{Conclusions and Further works} %

We have introduced the colored longest common prefix array (\clcp) that, given a strings collection $\bigS$, stores the length of the longest common prefix between the suffix of any string in $\bigS$ and the nearest suffix of another string of $\bigS$, according to the lexicographically sorted list of suffixes. This notion is extended in a natural way considering two collections of strings $\bigS^0$ and $\bigS^1$ in order to consider the longest common prefix between any pair of strings in different collections. 
We have also provided a lightweight method that computes $\clcp$ via sequential scans when $\bigS^0$ consists of a single string, but the strategy can be applied to the general case. This fact makes this data structure appropriate for computing several types of statistics on large collections of long strings. In particular, we have proved that $\clcp(\bigS)$ produces a permutation of the matching statistics (\MS) for the strings of the collection of $\bigS$. Based on this result, we have used $\clcp$ to efficiently solve the multi-string \ACS problem --- i.e.\ computing  pairwise MS between a string in $\bigS^0$ and all $m$ strings in $\bigS^1$ simultaneously, --- that is nowadays crucial in many practical applications, but demanding for large string comparisons.
This is also supported by experimental results. 

It is interesting to note that the data structure proposed in this paper, and its sequential strategy of computation, are intrinsically dynamic, i.e. $\clcp$ can be efficiently updated when the collection is modified by inserting or removing a string. In particular, after the removal of a string, $\clcp$ can be updated by exploiting the mathematical properties of the permutation associated with the $\EBWT$. The insertion of a new string in the collection can be managed by using merging strategy proposed in \cite{EgidiLouzaManziniTelles2018_arxiv} that works in semi-external memory. In this case the array $\arrayD$ can be constructed during this merging phase. 
Finally, we plan to extend our framework to solve the many-to-many pairwise \ACS problem on a collection $\bigS$ of $m$ sequences or between all strings of a collection versus all strings of another collection at the same time.  

\bibliographystyle{splncs04}
\bibliography{biblio} 

 \end{document}